\documentclass[10pt, conference]{IEEEtran}
\usepackage{cite}
\usepackage{amsmath,amssymb,amsfonts}
\usepackage{algorithmic}
\usepackage{graphicx}
\usepackage{textcomp}
\usepackage{xcolor}
\usepackage[hyphens]{url}

\usepackage{dcolumn}
\usepackage{bm}
\usepackage{enumerate}
\usepackage{mathtools}
\usepackage{physics} 
\usepackage{subcaption}
\usepackage{braket}
\usepackage[linesnumbered,ruled,vlined]{algorithm2e}
\DeclareMathOperator*{\argmax}{arg\,max}
\usepackage{enumitem}
\usepackage{multirow}

\def\BibTeX{{\rm B\kern-.05em{\sc i\kern-.025em b}\kern-.08em
    T\kern-.1667em\lower.7ex\hbox{E}\kern-.125emX}}

\begin{document}
\title{Divide and Conquer for Combinatorial Optimization and Distributed Quantum Computation} 


\author{
\IEEEauthorblockN{Teague Tomesh\IEEEauthorrefmark{1}\IEEEauthorrefmark{2}\IEEEauthorrefmark{4},
Zain H.~Saleem\IEEEauthorrefmark{3}\IEEEauthorrefmark{4},
Michael A.~Perlin\IEEEauthorrefmark{1},
Pranav Gokhale\IEEEauthorrefmark{1}, 
Martin Suchara\IEEEauthorrefmark{5} \\
and
Margaret Martonosi\IEEEauthorrefmark{2}} \IEEEauthorblockA{\IEEEauthorrefmark{1}Infleqtion, Chicago, IL, USA} \IEEEauthorblockA{\IEEEauthorrefmark{2}Department of Computer Science,
Princeton University, Princeton, NJ, USA}
\IEEEauthorblockA{\IEEEauthorrefmark{3}Mathematics and Computer Science Division,
Argonne National Laboratory, Lemont, IL, USA} \IEEEauthorblockA{\IEEEauthorrefmark{5}Amazon Web Services,
Amazon, Seattle, WA, USA}
\IEEEauthorblockA{\IEEEauthorrefmark{4}Email: teague.tomesh@infleqtion.com, zsaleem@anl.gov}
}




\maketitle
\begingroup\renewcommand\thefootnote{\textsection}
\footnotetext{The first two authors contributed equally to this work.}
\thispagestyle{plain}
\pagestyle{plain}


\begin{abstract}

Scaling the size of monolithic quantum computer systems is a difficult task. As the number of qubits within a device increases, a number of factors contribute to decreases in yield and performance. To meet this challenge, distributed architectures composed of many networked quantum computers have been proposed as a viable path to scalability.
Such systems will need algorithms and compilers that are tailored to their distributed architectures.
In this work we introduce the Quantum Divide and Conquer Algorithm (QDCA), a hybrid variational approach to mapping large combinatorial optimization problems onto distributed quantum architectures.
This is achieved through the combined use of graph partitioning and quantum circuit cutting. The QDCA, an example of application-compiler co-design, alters the structure of the variational ansatz to tame the exponential compilation overhead incurred by quantum circuit cutting.

The result of this cross-layer co-design is a highly flexible algorithm which can be tuned to the amount of classical or quantum computational resources that are available, and can be applied to both near- and long-term distributed quantum architectures.
We simulate the QDCA on instances of the Maximum Independent Set problem and find that it is able to outperform similar classical algorithms. We also evaluate an 8-qubit QDCA ansatz on a superconducting quantum computer and show that circuit cutting can help to mitigate the effects of noise. Our work demonstrates how many small-scale quantum computers can work together to solve problems $85\%$ larger than their own qubit count, motivating the development and potential of large-scale distributed quantum computing.

\end{abstract}


\section{Introduction}\label{sec:introduction}

Quantum computers (QCs) have the potential for computational speedups over classical computers for tasks related to cryptography \cite{shor1994algorithms}, simulating quantum systems \cite{lloyd1996universal, reiher2017elucidating}, and optimization \cite{farhi2014quantum, biamonte2017quantum}.
In order to realize an advantage, a QC must target large problem sizes beyond the reach of state-of-the-art classical algorithms. Therefore, the error rate during quantum program execution must be low enough to allow for successful program executions with thousands of qubits and millions of gates \cite{reiher2017elucidating, gidney2021factor}. 

Current Noisy Intermediate-Scale Quantum (NISQ) processors \cite{preskill2018quantum} have error rates which are too high (on the order of $0.05\%$ for single-qubit operations and $1\%$ for two-qubit operations \cite{tomesh2022supermarq}) and qubit counts which are too low (while most QCs currently support a few tens of qubits, a 127-qubit \cite{chow_dial_gambetta_2021} and a 433-qubit \cite{osprey_ibm_blog_2022} superconducting QC have also been built) to provide any practical advantages over classical algorithms.
To combat the effects of noise, quantum error correcting codes (QECCs) have been developed which reduce error rates by encoding a small number of high-fidelity, logical qubits using many noisy, physical qubits \cite{shor1995scheme, fowler2009high, nielsen_chuang_2010}.
The physical-to-logical encoding overhead incurred by QECCs further necessitates the need to scale the qubit counts of QC hardware.
In practice, scaling the size of monolithic QC systems is difficult because of adverse effects (e.g., frequency collision in superconducting systems \cite{ding2020crosstalk} or reduced gate speed and precision in trapped ion systems \cite{zhang2017observation, leung2018entangling}) which contribute to the deterioration of qubit coherence times, gate fidelities and processor yield rates \cite{smith2022scaling}.

Distributed QC architectures are a promising path towards achieving the scalability necessary for practically useful quantum computation \cite{cirac1999distributed, van2007communication, gyongyosi2021distributed}.  
The main tradeoff made by a distributed QC architecture appears in the tension between having smaller processors with more attractive noise characteristics and the need to perform remote operations between qubits on separate modules, which tend to be more expensive in terms of time, fidelity, or both when compared to intra-module operations \cite{van2007communication, gold2021entanglement, yan2022entanglement}.

In the long-term, distributed QC architectures will be supported by large-scale networks enabling quantum teleportation protocols to transfer quantum information between modules \cite{vanmeter2014internet}.
In the near-term, a distributed QC architecture may still be realized using recently introduced quantum circuit cutting techniques \cite{peng2020simulating}. 
Recent experiments have used many small quantum processing units (QPUs), combined with quantum circuit cutting, to simulate the execution of quantum programs which are larger than any one of the smaller QCs \cite{tang2021cutqc, perlin2021quantum}.
In both cases, distributed execution of quantum programs incurs an overhead cost that grows with the total number of remote operations.
The long-term the cost is measured by the number of entangled qubit pairs consumed, and in the near-term by the amount of classical post-processing that is required.

\begin{figure}[t]
    \centering
    \includegraphics[width=\columnwidth]{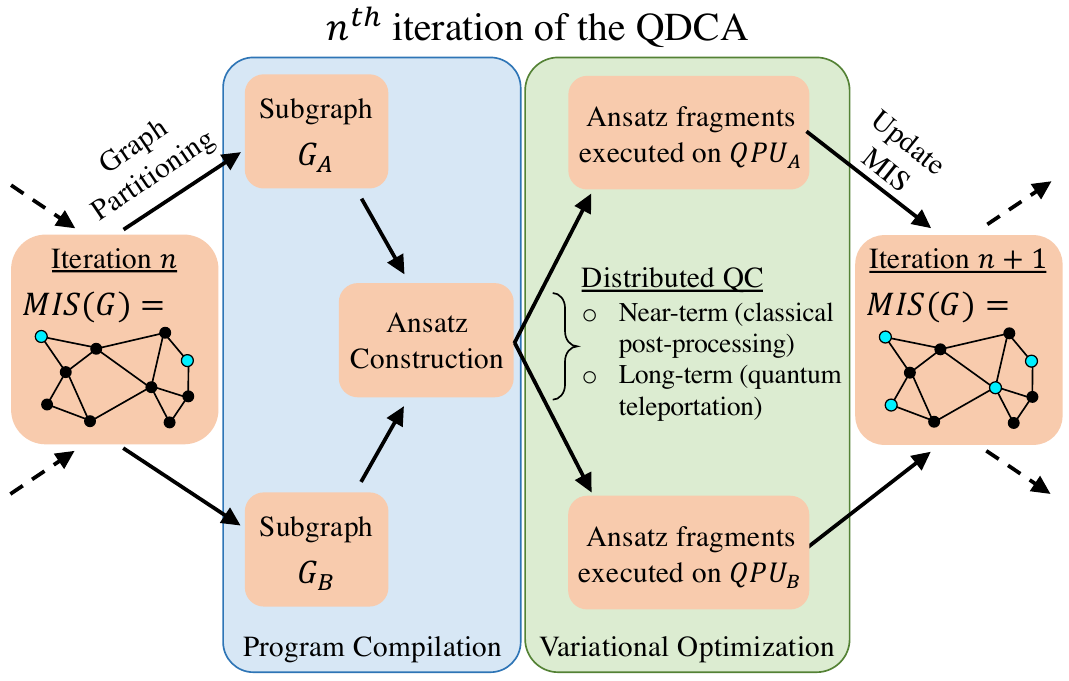}
    \caption{Overview of a single iteration within the QDCA. First, the target graph is partitioned into subgraphs. The subgraphs are used to construct a variational ansatz where a maximum of $m_{cuts}$ inter-subgraph quantum gates are allowed. After the ansatz is constructed, it is then executed on a distributed QC architecture (in the near-term this can be achieved via quantum circuit cutting techniques \cite{peng2020simulating, tang2021cutqc, perlin2021quantum}, while in the long-term this will be enabled by quantum teleportation protocols \cite{nielsen_chuang_2010, van2007communication}). Finally, the result of the variational optimization is used to update the current global solution.}
    \label{fig:overview}
\end{figure} 

New compilation tools and frameworks will be required to successfully navigate the tradeoffs between the remote operations and their associated overheads.
This work address this challenge by introducing the Quantum Divide and Conquer Algorithm (QDCA), a method for mapping large combinatorial optimization problems onto distributed QCs --- it shares many similarities with classical tools for distributing work among many nodes in high performance computing~\cite{chevalier2008pt}, or partitioning a computation too big to fit on a single FPGA~\cite{woo1993efficient}. At its core, the QDCA relies on variational (i.e., iterative) optimization to produce high quality solutions to the given combinatorial problem \cite{cerezo2021variational}. 

The QDCA relies on an application-compiler co-design approach \cite{tomesh2021codesign} to balance performance with the communication overhead incurred by distributed quantum computing. A high-level outline of the QDCA is shown in Figure \ref{fig:overview}.
We specifically target combinatorial optimization problems because they are amenable to divide and conquer strategies, allowing a problem on a large, target graph to be split into many, smaller subproblems.
After partitioning the target graph, a variational quantum circuit is constructed and cut. For example, in Figure \ref{fig:overview} the full ansatz contains gate operations acting on qubits in $QPU_A$ and $QPU_B$.
The circuit construction is carefully performed as to limit the communication overhead incurred by quantum circuit cutting.

In this work, we evaluate the performance of the QDCA via simulations and hardware evaluations and find that it is able to outperform similar classical algorithms while controlling the communication overhead in a tunable way.
Specifically, our contributions include:
\begin{itemize}
    \item A specification of the QDCA, including the partitioning of the input combinatorial optimization problem into multiple subproblems, the construction of the variational quantum circuit, and its execution on distributed QCs using quantum circuit cutting techniques.
    \item The first demonstration of quantum circuit cutting employed as a compilation tool within a hybrid, variational application. Using this technique we are able to find approximate solutions to Maximum Independent Set (MIS) problems on graphs containing up to 26 nodes while only requiring quantum circuits with up to 14 qubits.
    \item Finally, we evalute an 8-qubit QDCA ansatz on the superconducting \texttt{ibm\_algiers} QPU and find that the use of circuit cutting reduces the effects of noise.
\end{itemize}

The rest of the paper is organized as follows. Section \ref{sec:background} introduces the MIS problem, variational optimization algorithms, and quantum circuit cutting.
In Section \ref{sec:QDC} we provide a specification of the QDCA followed by a discussion in Section \ref{sec:considerations} of the architectural tradeoffs in both the near- and long-term.
We present our simulations and hardware evaluations of the QDCA in Section~\ref{sec:results} and discuss its relation to prior work in Section \ref{sec:priorwork}. Finally, in Section \ref{sec:conclusions} we conclude and discuss opportunities for future work.

\section{Background}\label{sec:background}
\subsection{The Maximum Independent Set Problem }\label{sec:classical}
The Maximum Indepedent Set (MIS) problem is an NP-Complete combinatorial optimization problem defined on a graph $G=(V, E)$ containing a set of nodes $V$ and edges $E$ \cite{karp1972reducibility, tarjan1977finding}. The goal is to find the largest possible independent set of $G$, where an independent set is a subset $S \subset V$ of nodes such that none of the nodes in $S$ are neighbors of one another.

In this work we focus on solving MIS instances using hybrid quantum-classical algorithms. Since these algorithms output binary bitstrings, we choose to define MIS using this notation. Given a graph $G$, with $N=\abs{V}$ nodes, we define a labelling of $G$'s nodes $V = \{v_i\}$ where $i \in [0, \dots, N-1]$, and each node's membership in the independent set is given by an $N$-bit bitstring $\bm{b} \in \{0,1\}^N$. For any bit $b_i \in \bm{b}$ if $b_i = 1$ then we say that node $v_i$ is a member of the independent set. Defining the MIS problem in this way means that only a subset of all possible $N$-bit strings correspond to valid independent sets. When we map the MIS problem onto a quantum computer we will assume the identity mapping $v_i \to q_i$ between graph vertices and qubits.

\subsection{Hybrid Variational Optimization for MIS}\label{subsec:optimization}

In a hybrid variational algorithm, a quantum and classical processor work together to optimize the expectation value of the problem's objective function. The classical computer runs an optimization routine (e.g., gradient descent),  iteratively updating a set of parameters that define a parameterized quantum circuit, called an \textit{ansatz}. During the optimization procedure, the objective function is evaluated using a QC by first preparing the quantum state defined by the current parameter values and then measuring the expectation value of this state with respect to the objective function.

As a concrete example, consider the MIS problem where we wish to maximize the objective function $C(\bm{b}) = \sum_{j \in V} b_j$, i.e., the Hamming weight of the solution bitstring $\bm{b}$. We convert the objective function into a quantum operator $C_{obj}$ which acts on quantum states $\ket{\bm{b}} = \ket{b_0 b_1 \dots b_{n-1}}$ as
\begin{equation} \label{qoperator}
    C_{obj} \ket{\bm{b}} = C(\bm{b}) \ket{\bm{b}} = \sum_{j \in V} \frac{1 - Z_j}{2} \ket{\bm{b}}, 
\end{equation}
where $Z_j$ is the Pauli-$Z$ matrix $\begin{pmatrix} 1 & 0 \\ 0 & -1\end{pmatrix}$ acting on qubit $j$.

Given a list of parameters $\bm{\theta}$, where $\theta_i \in \mathbb{R}$, the QC computes value of the objective function by measuring the expectation
\begin{equation}
    E(\bm{\theta})
    = \braket{\psi \left(\boldsymbol{\theta}\right) | C_{obj}
    |\psi \left(\boldsymbol{\theta}\right)}. 
\end{equation}
The state $\ket{\psi(\bm{\theta})}$ is referred to as the variational ansatz.

The Quantum Approximate Optimization Algorithm (QAOA) \cite{farhi2014quantum} defines a broad class of algorithms, which includes the QDCA, for solving combinatorial optimization problems. In the QAOA, the ansatz is typically composed of three main parts: an easy to prepare initial state $\ket{s}$, the phase separation unitary $e^{-i \gamma C}$, and the mixing unitary $e^{-i \alpha M}$, which are applied in $p$ alternating layers:
\begin{equation}\label{eqn:expectation}
    \ket{\psi_p \left(\boldsymbol{\gamma},\boldsymbol{\alpha} \right)}
    = e^{-i\alpha_p M}e^{-i\gamma_p C}\dots e^{-i\alpha_1 M}e^{-i\gamma_1 C}\ket{s}.   
\end{equation}
Note that we have split the list of variational parameters $\bm{\theta} \coloneqq (\bm\gamma, \bm\alpha)$ into two separate lists for convenience.

Finally, we must consider how to impose the MIS constraints on the QAOA. One option would be to follow the original formulation of the QAOA which was given for unconstrained optimization problems such as MaxCut \cite{farhi2014quantum}. This approach can be used to solve \textit{constrained} problems, like MIS, by modifying the objective function to heavily penalize constraint-breaking bitstrings. This converts the constrained optimization into an unconstrained one, however, in this approach, one must now decide how large the penalty term should be and perform an additional post-processing step to prune any invalid bitstrings \cite{saleem2020approaches}.
In this paper, we instead choose to impose the MIS constraints within the variational ansatz itself. We refer to this approach as the constrained QAOA, and the idea is to modify the QAOA's mixing unitary to ensure that the quantum state always remains within the subspace of feasible solutions \cite{hadfield2018quantum, hadfield2019quantum, saleem2020max}.

\begin{figure}
\centering
\includegraphics[width=0.9\columnwidth]{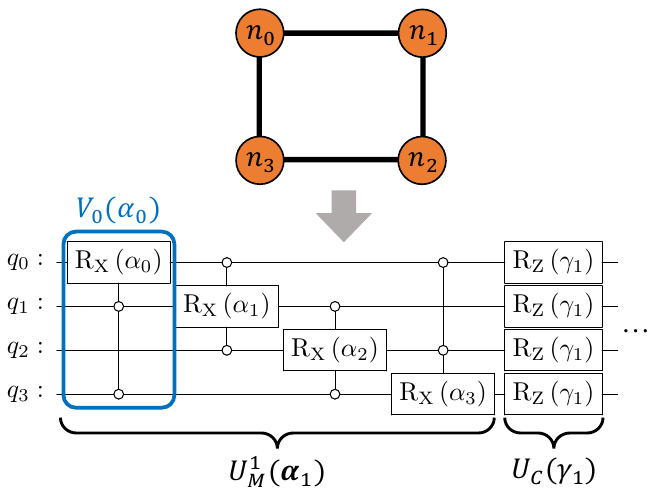}
\caption{An example graph (top) is mapped to a variational ansatz (bottom) where node $n_i$ corresponds to qubit $q_i$. Together, the partial mixers (blue box), which apply a conditional rotation controlled by a node's neighbors, form the mixing unitary $U^1_M(\bm\alpha_1)$. The phase separator unitary $U_C(\gamma_1)$ is composed of single-qubit Z-rotations.} 
\label{fig:ansatz_example}
\end{figure}

In the constrained QAOA, the initial state $\ket{s}$ is chosen to be a feasible state (e.g., the all-zero state) or a superposition of feasible states (e.g., a Dicke state which is an equal superposition over all unit Hamming weight bistrings \cite{dicke1954coherence}) corresponding to valid independent sets. The phase separation unitary $U_C(\gamma) = e^{-i\gamma C_{obj}}$ is constructed using the objective function operator defined in Equation \eqref{qoperator} and shown in Figure \ref{fig:ansatz_example}. Finally, the mixing unitary is constructed in such a way that a qubit can only rotate from $\ket{0}$ to $\ket{1}$ if none of its neighbors are in the state $\ket{1}$ (equivalently, if all of its neighbors are in $\ket{0}$).

Encoding the MIS constraints in this manner results in a mixing unitary of the form $U_{M}(\alpha) = e^{-i \alpha M} = \prod_j V_j(\alpha)$, where the partial mixers $V_j(\alpha)=e^{i \alpha M_j}$ correspond to conditional rotations on each qubit $j$ and are generated by $M_j = X_j \bar{B}_j$, where 
\begin{align}
\bar{B}_j =  \prod_{k\in\mathcal{N}(j)} \bar{b}_k,
&&
\bar{b}_k = \op{0}_k = \frac{1+Z_k}{2}.
\end{align}
Here, $X_j$ is the Pauli-$X$ matrix $\begin{pmatrix} 0 & 1 \\ 1 & 0\end{pmatrix}$ acting on qubit $j$, and $\mathcal{N}(j)$ denotes the neighbors of node $j$.
Using the fact that the projector $\bar B_j^2=\bar B_j$, the partial mixers can equivalently be written as
\begin{align}
V_j(\alpha) = \bar B_j e^{i\alpha X_j} + (1-\bar B_j) I,
\end{align}
where $I$ is the identity operator, this emphasizes the interpretation of these partial mixers multi-controlled qubit rotations as shown in Figure \ref{fig:ansatz_example}.

As the partial mixers $V_j$ generally do not commute with each other ($V_j V_k \neq V_k V_j$), it is more natural to define a mixing unitary equipped with some permutation $\sigma$ of $\{1,2,\cdots,n\}$:
\begin{align}
\label{eqn:qao-mixer}
U_M^\sigma(\alpha)
= V_{\sigma(n)}(\alpha)\cdots V_{\sigma(2)}(\alpha) V_{\sigma(1)}(\alpha).
\end{align}

Finally, instead of parameterizing each partial mixer $V_j(\alpha)$ with the same angle, we allow the mixing unitary to take a list of angles as input $U_M(\alpha) \rightarrow U_M(\bm\alpha)$ such that each partial mixer is individually parameterized.
This provides the classical optimizer with fine-grained control over the qubit rotations, and, in practice, this has been shown to lead to improved performance with smaller circuit depth \cite{herrman2022multi}. In addition, uniquely parameterized partial mixers enables some of the partial mixers to be turned off by setting $\alpha_i = 0$, effectively reducing the gate cost of the ansatz, which may be beneficial for hardware implementations \cite{saleem2020approaches}.

\subsection{Distributed Quantum Computing with Circuit Cutting}\label{sec:distributedQC}
\begin{figure}[t]
\centering
\includegraphics[width=0.7\columnwidth]{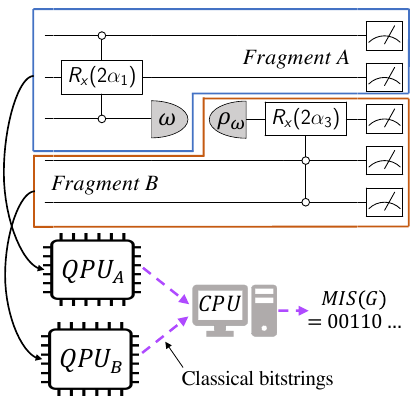}
\caption{Example of quantum circuit cutting to execute a program on a distributed QC. Quantum circuit cutting techniques \cite{peng2020simulating, perlin2021quantum, tang2021cutqc} can be used to split a large circuit into smaller fragments. These fragments are then executed independently on separate QPUs, and the output of the original circuit is reconstructed via an exponentially costly classical post-processing process.
}
\label{fig:distributedQC}
\end{figure}

In the current NISQ era and into the foreseeable future, qubits will remain a scarce resource. For variational algorithms such as the QAOA, scaling to large problem sizes is limited by the qubit count of the underlying hardware. This is the result of 1:1 mappings between problem variables and qubits (as was done for the MIS problems in Section \ref{subsec:optimization}), and this issue becomes even more severe when larger constant or linear overheads are needed to map a problem onto a QPU --- for example, when encoding an asset position into the state of two qubits in the context of portfolio optimization \cite{slate2021quantum}.

One solution is to simply increase the number of qubits on a monolithic QPU, but as more qubits are added to a single device, the yield or overall performance tends to decrease. This can be attributed to a number of factors specific to certain qubit technologies. Some examples include frequency collisions~\cite{ding2020crosstalk}, crowded control wiring, and increased ion chain fluctuations~\cite{murali2020architecting}.

A distributed QC architecture circumvents the scalability issues associated with monolithic architectures by repeatedly manufacturing many fixed-sized units and networking them together~\cite{cirac1999distributed, van2007communication, gyongyosi2021distributed}. 
In the current NISQ era, many dozens of small-scale QCs have been manufactured and made available by various cloud providers \cite{cross2018ibm, braket2020}. Typically, these processors have been considered as standalone systems, but recent work has introduced a new technique which allows many independent QPUs to work together to emulate the execution of a larger, monolithic QC \cite{peng2020simulating, tang2021cutqc, perlin2021quantum}.

This technique, known as \textit{quantum circuit cutting}, cuts a quantum circuit (i.e., a quantum program) into multiple subcircuits which can be independently executed, and their results ``stitched together'' to reconstruct the output of the full circuit. This is done by correlating measurement outcomes, $\omega$, on the ``output'' of one subcircuit (at a cut location) with initial states, $\rho_\omega$, prepared at the ``input'' of another subcircuit, see the quantum circuit shown in Figure \ref{fig:distributedQC}.

Importantly, as more cuts are made to the original circuit, the overhead cost of the classical reconstruction step increases \textit{exponentially} \cite{peng2020simulating, perlin2021quantum, tang2021cutqc}. Consider a single circuit fragment containing $K_i$ cut inputs (i.e., the $\rho_\omega$ state in Fragment $B$ of Figure~\ref{fig:distributedQC}) and $K_o$ cut outputs (i.e., the $\omega$ measurements in Fragment $A$ of Figure~\ref{fig:distributedQC}). Reconstructing the output of the original uncut circuit requires evaluating many different variants of its circuit fragments, with different combinations of state preparations on the $K_i$ cut inputs and different measurements at the $K_o$ cut outputs of any given fragment. Specifically, if a circuit is cut into fragments using $k$ cuts, then reconstruction requires preparing and evaluating $\mathcal{O}(4^{K_i}3^{K_o})$ subcircuit variants, and then obtaining the output of the original circuit by performing $\mathcal{O}(4^k)$ Kronecker products on a classical computer.

In the context of variational algorithms, the circuit cutting technique is used to evaluate the objective function while using fewer qubits than would otherwise be required. To obtain an estimate of the objective function to precision $\epsilon$, $O(\sum_\mu 4^{K^\mu_i}3^{K^\mu_o} / \epsilon^2)$ shots must be allocated on the quantum computer, assuming that the shots are distributed evenly across all circuit fragments (indexed by $\mu$). However, this is a conservative assumption as there are many prior works which have investigated more optimal shot allocation strategies for circuit cutting~\cite{tang2021cutqc} and algorithms such as VQE~\cite{kohda2021quantum, gokhale2020n}. 

Reducing the overhead costs of circuit cutting is a major concern for any practical implementation. This is why the QDCA was developed specifically to control the number of cuts that need to be applied to split the variational ansatz. In Section~\ref{sec:QDC} we describe how the post-processing overhead is controlled via a user-determined hyperparameter that can be tuned to match the available classical post-processing resources.

\begin{algorithm}[t]
\SetAlgoLined
\SetKwData{Left}{left}\SetKwData{This}{this}\SetKwData{Up}{up}
\SetKwFunction{Union}{Union}\SetKwFunction{FindCompress}{FindCompress}
\SetKwInOut{Input}{Input}\SetKwInOut{Output}{Output}

\Input{$G = (Q,E)$, $N$ = \# of partition rounds, \\ $m$ = max \# of cuts}
\Output{Approximate MIS of $G$}
$s \leftarrow$ initial state selection\;
$s_{best} \leftarrow ``00...0"$\;
\For{$r \in [N]$}{
$h_{new}, h_{old} \leftarrow H(\text{s}), -1$\;
\tcc{Classical Graph Partitioning}
$G_A, G_B, Q^c \leftarrow \text{bisection(G)}$\;
\While{$h_{new} > h_{old}$}{
  \tcc{Hot Node Selection}
  $Q^{hot} \leftarrow \text{min}(\mathcal{S}(Q^{hot}), \text{ s.t. } \abs{\mathcal{I}(Q^{\mathrm{hot}})} \leq m)$\;
  \tcc{Ansatz Construction}
  $\Vec{\alpha} \leftarrow [0_1, 0_2, \dots, 0_n]$\;
  $\gamma \leftarrow \text{random}(0,2\pi)$\;
  \For{$q \in \{Q^{uc}_A \bigcup Q^{uc}_B \bigcup Q^{hot}\}$}{
    $\alpha_q \leftarrow \text{random}(0,2\pi)$\;
  }
  $U_{ansatz}(\Vec{\alpha}, \gamma) \leftarrow U_M(\Vec{\alpha})U_C(\gamma)$\;
  \tcc{Variational Optimization}
  \While{not converged}{
    \tcc{Quantum Circuit Cutting}
    subcircs, cuts $\leftarrow$ cut($U_{ansatz}(\Vec{\alpha}, \gamma)\ket{s}, m$)\;
    subcirc\_counts $\leftarrow$ evaluate(subcircs)\;
    counts $\leftarrow$ reconstruct(subcirc\_counts, cuts)\;
    $E \leftarrow$\ expectation\_value($H$, counts)\;
  }
  
  $h_{old} \leftarrow H(s)$\;
  $h_{new} \leftarrow \max_b([H(b) \text{ for } b \text{ in counts}])$\;
  $s \leftarrow \argmax_b{([H(b) \text{ for } b \text{ in counts}])}$\;
  \If{$h_{new} > H(s_{best})$}{
    $s_{best} \leftarrow s$\;
  }
}
}
 \KwRet $s_{best}$
 \caption{QDCA (Classical-Only Communication)}
 \label{alg:qdca}
\end{algorithm}

\section{The Quantum Divide and Conquer algorithm}\label{sec:QDC}

In this section, we define each component of the QDCA following the pseudocode in Algorithm \ref{alg:qdca}. An open-source implementation of the QDCA is available via GitHub \cite{dqva2023github}.

A high-level outline of the QDCA is shown in Figure \ref{fig:overview} where, in each iteration, the target graph is partitioned into smaller subgraphs and then a variational ansatz is constructed over these subgraphs.
It is important to distinguish between the two levels of partitioning (cutting) which take place within the QDCA. First, the input graph must be \textit{partitioned} into two separate subgraphs. Then the quantum circuit that is constructed is \textit{cut} into two subcircuits which are executed on a distributed QC architecture. For clarity, we try to reserve the term ``partition'' for referring to the input graph, and ``cut'' for referring to the quantum circuit. Note that the number of split edges in the graph partition may not equal the number of cut qubits. 



\subsection{Classical Graph Partitioning}
Given a graph $G=(Q,E)$, partition the graph into two subgraphs $G_A$ and $G_B$ of roughly equal sizes using a partitioning algorithm to minimize the number of crossing edges; in this work we consider the classical Kernighan-Lin and METIS algorithms~\cite{kernighan1970efficient, karypis1997metis}.

After the graph is partitioned, each node will belong to one of four sets. The nodes in $G_A$ and $G_B$ with no edges crossing the partition are members of the sets $Q^{uc}_A$ and $Q^{uc}_B$, respectively. Whereas, those nodes with at least one edge crossing the partition belong to the sets $Q^c_A$ and $Q^c_B$.

In subsequent iterations of the QDCA, the graph partitioning step may be biased to favor partitioning edges containing qubits that have been assigned to $\ket{1}$ in the initial state (i.e.~included in the current independent set).
This is preferable because these vertices do not require any partition-crossing partial mixers (discussed in the steps below).
The biasing may be accomplished by reducing the weight of edges involving vertices which have been assigned to $\ket{1}$ in the initial state.
Thereafter, the graph partitioning algorithm should minimize the weight (rather than the number) of partition-crossing edges.
    
\subsection{Hot Node Selection}
\begin{figure}[t]
    \centering
    \includegraphics[width=0.9\columnwidth]{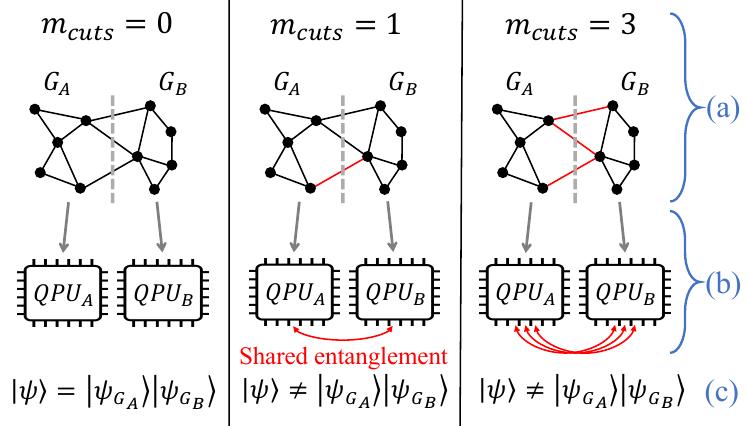}
    \caption{Three possible QDCA compilations showing how the same graph partition (dashed grey line) can lead to (a) more cross-partition quantum gates (red lines) with increasing values of $m_{cuts}$. This corresponds to (b) more shared entanglement between distributed QPUs, and when (c) the ansatz is evaluated the final output state $\ket{\psi}$ may be entangled if $m_{cuts}>0$.}
    \label{fig:regimes}
\end{figure}
We now need to select the set $Q^{\mathrm{hot}}\subset Q^c$ of ``hot nodes''. This is the subset of partitioned nodes $Q^c\equiv Q_A^c\cup Q_B^c$ to which we will apply nontrivial partial mixers that span the two subgraphs.
In contrast, the ``cold nodes'' $Q^{\mathrm{cold}}=Q^c\setminus Q^{\mathrm{hot}}=\{q\in Q^c:q\notin Q^{\mathrm{hot}}\}$ will have their partial mixers ``turned off'' by setting the rotation angle $\alpha_q=0$ for all nodes $q\in Q^{\mathrm{cold}}$. With $\alpha_q=0$, the partial mixer is a trivial gate equivalent to the identity operation.

As the number of cut nodes $\abs{Q^c}$ is typically small, we use a brute-force search to find the set of hot nodes $Q^{\mathrm{hot}}$ that maximizes the score function
\begin{align}
\mathcal{S}(Q^{\mathrm{hot}})
= \sum_{i\in Q^{\mathrm{hot}}}
\sum_{\substack{j\in \mathcal{N}(i)\\g(j)\ne g(i)}}
\mathcal{D}(j),
\end{align}
where $g(q)\in\{G_A,G_B\}$ is the subgraph containing node $q$, and $\mathcal{D}(q)$ is the degree of node $q$.

The maximization of $\mathcal{S}(Q^{\mathrm{hot}})$ is performed under the constraint that the number of cross-partition neighbors of the hot nodes, i.e.~the size of the set
\begin{align}
\mathcal{I}(Q^{\mathrm{hot}})
= \bigcup_{i\in Q^{\mathrm{hot}}}
\{ j\in\mathcal{N}(i) : g(j) \ne g(i) \},
\end{align}
does not exceed some number $\lvert \mathcal{I}(Q^{\mathrm{hot}}) \rvert \leq m_{cuts}$ of maximal cuts that will be tolerated for circuit cutting.
This score function heuristically tries to maximize the amount of inter-subgraph entanglement.
The effect of the $m_{cuts}$ hyperparameter is shown in Figure \ref{fig:regimes} where more entanglement (red edges) is generated across the graph partition (the dashed grey line) as $m_{cuts}$ increases.

Entanglement between subgraphs is the key factor which enables the QDCA to find larger independent sets than an uncorrelated classical divide and conquer approach. In the limit where $m_{cuts}$ is large enough to allow all of the partitioned edges to be included in the ansatz this is equivalent to performing QAOA over the full graph. However, entanglement between subgraphs comes at the cost of the increased overhead required to implement the remote gates.

In principle, the number of cuts required to separate the variational ansatz depends on the choice of partial mixer order.
To simplify the brute-force search for hot nodes, we therefore enforce that partial mixers are always ordered by their subgraph (we test both possible orders), and that the set of hot nodes is chosen from the first subgraph in this ordering.
This restriction also ensures that the number of circuit cuts that will be required for our ansatz is precisely $\abs{\mathcal{I}(Q^{\mathrm{hot}})}$.

\subsection{Ansatz Construction}
For a chosen integer $p$, create the variational ansatz 
\begin{equation}
    U_\text{ansatz} = U_C(\gamma_p) U_M(\boldsymbol\alpha_p) \cdots U_C(\gamma_1) U_M(\boldsymbol\alpha_1)
\end{equation} 
with some (fixed) order of partial mixers.
In the $k=1$ layer, fix $\alpha_1^j=0$ for all cold nodes $j\in Q^{\mathrm{cold}}$, and subsequently fix $\alpha_k^j=0$ in all layers $k>1$ for all cut nodes $j\in Q^c$.

The initial state $\ket{s}$ should also be set to the best solution that has been found thus far by applying an $X$ gate to each qubit in the independent set.

\subsection{Variational Optimization}
Once $U_{ansatz}$ has been constructed, optimize its free parameters within a quantum-classical variational loop. After measuring the output of the optimized circuit, store the highest Hamming weight bitstring which is also a valid independent set.

The quantum circuit cutting technique is applied within this loop to enable distributed program executions, effectively increasing the size of the quantum circuits that can be run on the quantum computer and therefore allowing the QPUs to target larger graphs~\cite{perlin2021quantum, tang2021cutqc}. 

Finally, the above steps may be repeated until a maximum number of partition rounds is reached or the optimization converges to a solution. During each partition round, a new random bi-partition of the graph is found, and the initial state is set to the best feasible state (with largest Hamming weight) observed in the prior round.

\section{Architectural Considerations of the QDCA}\label{sec:considerations}
The main advantage of the QDCA compared to classical divide and conquer algorithms is its ability to share information between the subproblems produced after the target graph has been partitioned. Rather than treating the subproblems as entirely independent, the QDCA constructs a quantum state across the subproblems by allowing a carefully controlled number of partial mixers to cross the partition (see Figure \ref{fig:regimes}). 

As discussed in Section~\ref{sec:distributedQC} the cross-partition gates can be implemented through circuit cutting (i.e., reconstructing the entanglement via classical post-processing), and in the future they may be implemented via quantum teleportation by coherently transferring quantum information between QPUs.

The communication overhead incurred by the teleportation strategy is straightforward to compute. For a circuit which can be split into two subcircuits using $k$ cuts, then $k$ teleportations could be inserted into the program at the cut points to move the necessary qubits between QPUs. This strategy avoids the need for an exponential amount of classical post-processing, and the latency overhead could be minimized if a sufficient number of entangled Bell pairs were generated and distributed ahead of the computation.

In the case of quantum circuit cutting, consider the case where $k$ cuts are made to split the original circuit into only two subcircuits. Since the cut qubits will only appear once as either an input or output qubit for each subcircuit, the total number of fragments is bounded by $O(2^{2k+1})$. Combining this with the classical post-processing that must be done gives an overall overhead of $O(4^k(2t_q + t_p))$ where $t_q$ is the time to evaluate a single quantum circuit and $t_p$ is the time to evaluate a single Kronecker product. It is reasonable to assume that because the gate times of current NISQ processors are on the order of microseconds~\cite{tomesh2022supermarq}, and the classical Kronecker products can be performed in parallel on GPUs with speeds on the order of gigaflops~\cite{tang2020high}, that the $t_q$ term will dominate this expression. Then, an overall runtime of $O(2^{2k+1}t_q)$ is necessary to reconstruct the output distribution of the original circuit. In addition to the parallelizable subcircuit executions and classical reconstruction operations, recent works have proposed numerous techniques to mitigate this overhead cost by discarding irrelevant portions of the computation \cite{tang2021cutqc, chen2023efficient}.

\begin{figure*}[th!]
\centering
    \begin{subfigure}{\textwidth}
    \includegraphics[width=\linewidth]{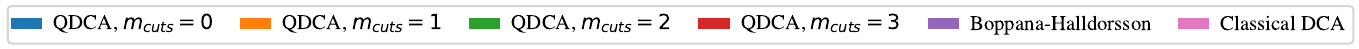}
    \end{subfigure}
    \hfill
    \begin{subfigure}{\columnwidth}
    \includegraphics[width=\columnwidth]{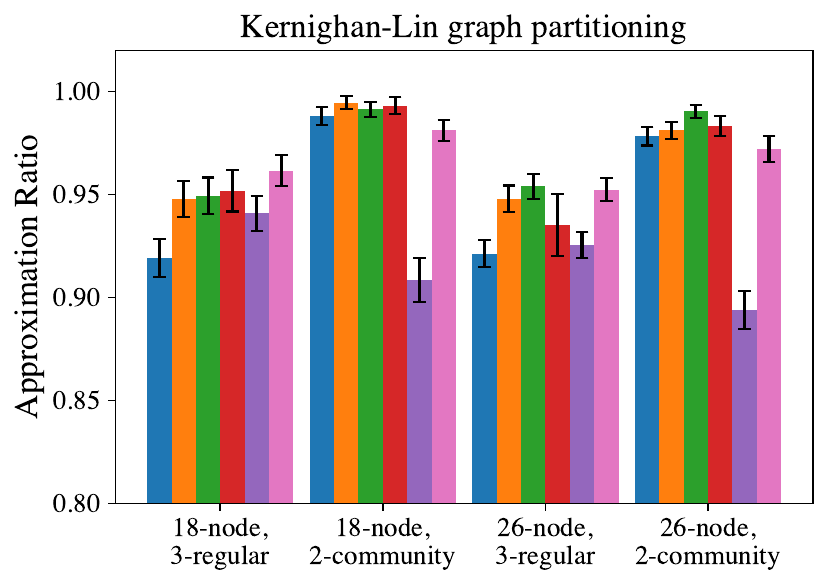}
    \caption{}
    \label{subfig:klb-barplot}
    \end{subfigure}
    \hfill
    \begin{subfigure}{\columnwidth}
    \includegraphics[width=\columnwidth]{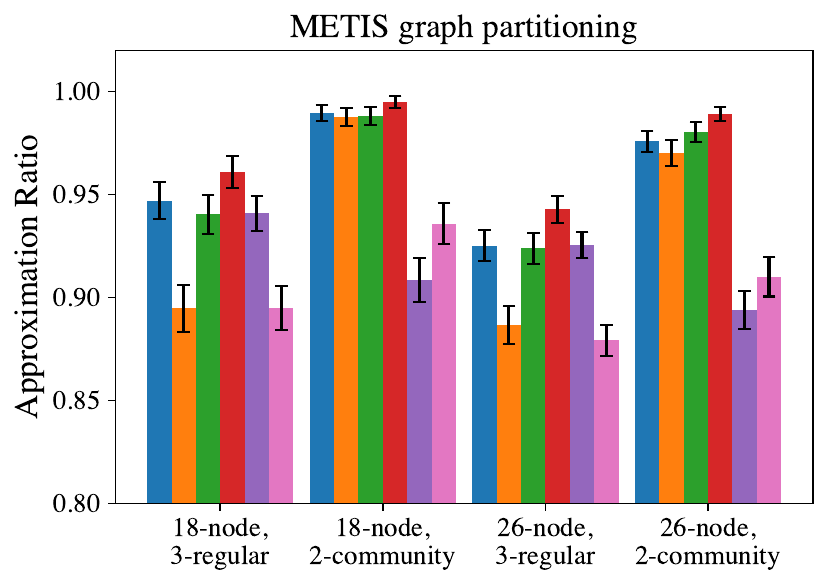}
    \caption{}
    \label{subfig:metis-barplot}
    \end{subfigure}
\caption{Average final approximation ratio of the QDCA (with varying values of $m_{cuts}$), CDCA, and Boppana-Halldorsson when solving MIS instances over 60 benchmark graphs of varying types of 18 and 26 nodes. The error bars denote the standard error from the mean.}
\label{fig:barplots}
\end{figure*}

\begin{figure*}[t]
\centering
    \begin{subfigure}{\textwidth}
    \includegraphics[width=\linewidth]{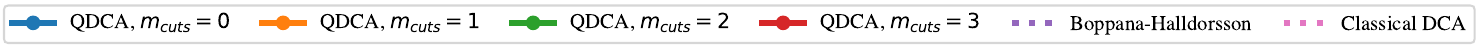}
    \end{subfigure}
    \hfill
    \begin{subfigure}{\columnwidth}
    \includegraphics[width=\columnwidth]{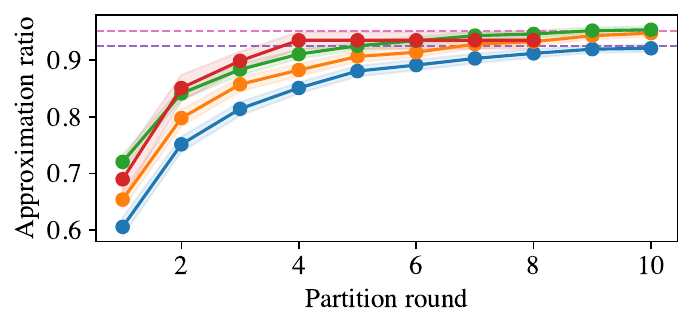}
    \caption{26-node, 3-regular graphs, KLB partition}
    \label{subfig:26nodeD3klb}
    \end{subfigure}
    \hfill
    \begin{subfigure}{\columnwidth}
    \includegraphics[width=\columnwidth]{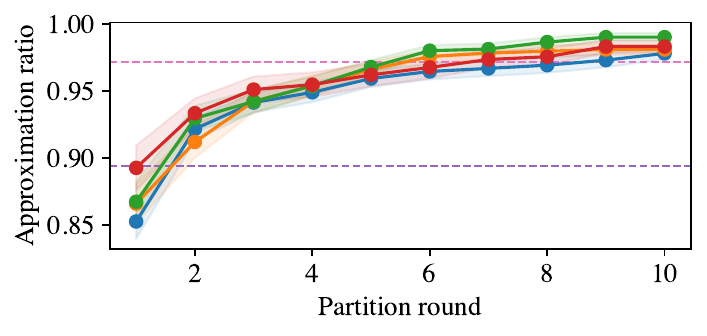}
    \caption{26-node, 2-community graphs, KLB partition}
    \label{subfig:26nodeCom2klb}
    \end{subfigure}
    \hfill
    \begin{subfigure}{\columnwidth}
    \includegraphics[width=\columnwidth]{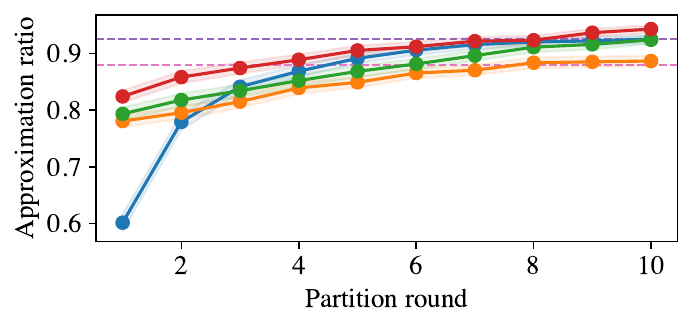}
    \caption{26-node, 3-regular graphs, METIS partition}
    \label{subfig:26nodeD3metis}
    \end{subfigure}
    \hfill
    \begin{subfigure}{\columnwidth}
    \includegraphics[width=\columnwidth]{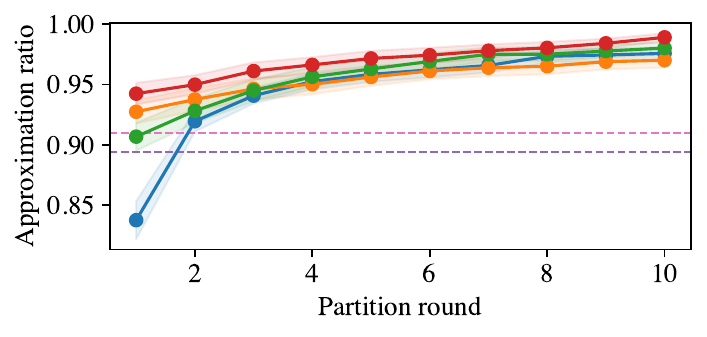}
    \caption{26-node, 2-community graphs, METIS partition}
    \label{subfig:26nodeCom2metis}
    \end{subfigure}
\caption{performance over iterations.}
\label{fig:iterations}
\end{figure*}

\section{Simulation and Hardware Results} \label{sec:results}
In Section \ref{subsec:noiseless}, we evaluate the performance of the QDCA, using noiseless simulation, on both 3-regular and 2-community graphs containing $N=18 \text{ and } 26$ nodes. The 3-regular graphs are created with the NetworkX \texttt{random\_regular\_graph(d=3, n=N)} function; in these graphs each node is connected to exactly three neighbors~\cite{hagberg2008exploring}. The 2-community graphs are created with the \texttt{planted\_partition\_graph(L=2, k=int(N/2), p\_in=0.2, p\_out=0.02)} NetworkX function. These random graphs are composed of $\texttt{L}=2$ communities with $\texttt{k}=\texttt{int}(N/2)$ nodes in each community. Edges between nodes in the same community exist with probability $\texttt{p\_in} = 0.2$ while edges between communities exist with probability $\texttt{p\_out} = 0.02$~\cite{condon2001algorithms}. The QDCA is implemented using Qiskit~\cite{cross2018ibm}, the COBYLA optimizer available in SciPy \cite{2020SciPy-NMeth}, and the circuit cutting software made available in~\cite{perlin2021quantum}. 

In Section \ref{subsec:hardware} we investigate the impact of noise on the QDCA execution. For an 8-node example graph, we compare the output distributions produced by ideal and noisy simulations as well as execution on the superconducting quantum computer \texttt{ibm\_algiers} --- with and without circuit cutting.

Our implementation is available online in an open source GitHub repository~\cite{dqva2023github}. It contains the QDCA as well as the classical algorithms used for comparison including Boppana-Halld\'orsson~\cite{boppana1992approximating} -- a quadratic runtime, recursive algorithm -- and a classical divide and conquer algorithm (CDCA) which uses Boppana-Halld\'orsson as a subroutine. The Boppana-Halld\'orsson algorithm is applied to the entire problem graph at once while the CDCA recursively partitions the graph and solves the resulting subproblems using Boppana-Halld\'orsson. 

\subsection{Noiseless simulation results}\label{subsec:noiseless}

Figure~\ref{fig:barplots} shows the average final approximation ratio obtained by the QDCA, the CDCA, and Boppana-Halld\'orsson on the different sets of benchmark graphs partitioned using both the Kernighan-Lin algorithm~\cite{kernighan1970efficient}, and the METIS graph paritioning algorithm~\cite{karypis1997metis}.
The approximation ratio $\left| S \right| / \left| S^* \right|$ is defined by the ratio between the size of the independent set $S$ found by a given algorithm and the optimal, maximum independent set $S^*$ found via brute force search.

The performance of the QDCA remains relatively constant regardless of the choice of graph partitioner, while the performance of the CDCA and Boppana-Halld\'orrson is best when using the Kernighan-Lin algorithm. 
For the graph partitioned using Kernighan-Lin (Figure \ref{subfig:klb-barplot}) the performance of the QDCA tends to improve as the value of $m_{cuts}$ increases, although for most of the graphs, $m_{cuts} = 1,2,3$ achieve approximately the same approximation ratios. The main difference between these runs appears in their time to convergence, which can be seen clearly in Figure \ref{fig:iterations}, where the larger values of $m_{cuts}$ rapidly improve the average size of the independent set as a function of partition rounds.

When the graphs are partitioned with the METIS algorithm, there is a larger difference in the final approximation ratios of the QDCA. The $m_{cuts}=0$ base case actually performs quite well, matching the average performance of the Boppana-Halld\'orsson algorithm for the 3-regular graphs. The $m_{cuts}=0$ runs are able to find a good local-minimum which is only surpassed once $m_{cuts}=3$.

\begin{figure}
    \centering
    \includegraphics[width=\columnwidth]{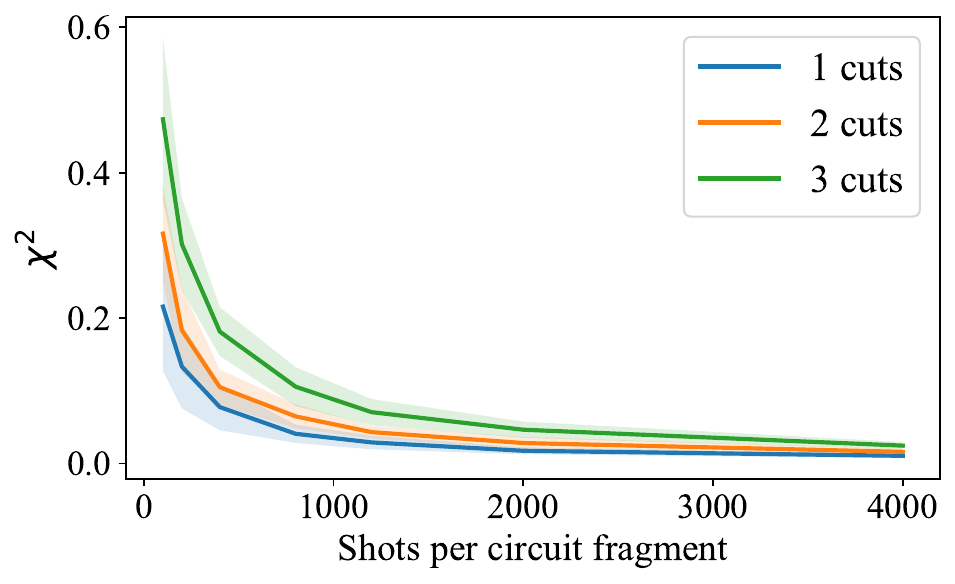}
    \caption{The average $\chi^2$ loss (lower is better) between 18-qubit cut and uncut QDCA circuits as a function of the number of samples obtained from each circuit fragment. The reported value is averaged over ten independent trials using random circuit parameters, the shaded regions show one standard deviation from the mean.}
    \label{fig:chisquared}
\end{figure}

Quantum circuit cutting techniques are known to introduce shot noise into the final output distribution when recombining the outputs of the finitely sampled subcircuits \cite{perlin2021quantum}.
Figure \ref{fig:chisquared} shows the impact of the sampling noise on the reconstructed output distribution when splitting an 18-qubit QDCA circuit using 1, 2, and 3 cuts. We compare the reconstructed output distribution to the ground truth (i.e., uncut) distribution using the $\chi^2$ loss, as proposed in prior circuit cutting work \cite{tang2021cutqc}.
\begin{equation}
    \chi^2 = \sum^{2^n-1}_{i=0}\frac{(p_i - q_i)^2}{p_i + q_i}
\end{equation}
Here, $p_i$ and $q_i$ are elements from the probability distributions obtained from the cut and uncut $n$-qubit circuits. For a given number of shots sampled from each circuit fragment, Figure \ref{fig:chisquared} reports the average $\chi^2$ value found over ten trials; in each trial the 18-qubit QDCA circuit is parameterized by a new set of random rotation angles.
In order to mitigate the effects of this fundamental shot noise, the results shown in Figures \ref{fig:barplots} and \ref{fig:iterations} were obtained using more than 5000 shots per circuit fragment.

\begin{figure}
    \centering
    \includegraphics[width=0.85\columnwidth]{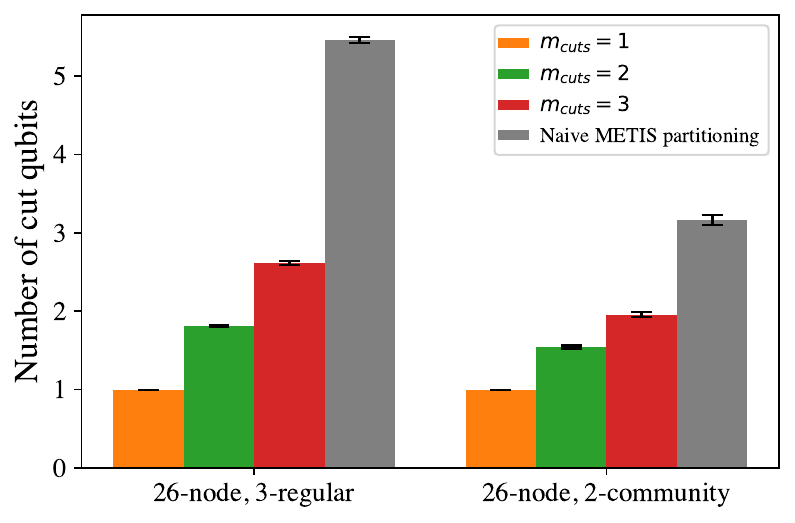}
    \caption{Average number of cut qubits in the QDCA circuits when using the METIS graph partitioner. For circuit cutting, the classical post-processing cost scales \textit{exponentially} with the number of cut qubits. The QDCA controls this overhead via the $m_{cuts}$ parameter.}
    \label{fig:cutqubits}
\end{figure}

A key aspect of the QDCA is its ability to control the amount of communication between the distributed QPUs when executing the variational ansatz. In the case of quantum circuit cutting this appears as the number of cuts required to split the variational ansatz. Figure \ref{fig:cutqubits} shows the average number of cut qubits needed for the QDCA with $m_{cuts}=1,2,3$, note that the average value tends to be less than $m_{cuts}$ because this hyperparameter only sets a limit to the number of cuts --- some of the ansatzes can be split using fewer than the maximum allowed cuts. The number of cuts required by the QDCA is compared with the number of cuts that would be required when partitioning the graph with the METIS algorithm and constructing the ansatz over the full graph (grey bars in Figure \ref{fig:cutqubits}). 

\begin{table}[]
\resizebox{\columnwidth}{!}{%
\begin{tabular}{cccc}
\hline
\multirow{2}{*}{Graph} & \multicolumn{3}{c}{(Avg \# cuts, Avg \# qubits, Max \# qubits)} \\
                       & $m_{cuts} = 1$         & $m_{cuts} = 2$         & $m_{cuts} = 3$         \\ \hline
N18 3-reg                 & 1.0, 10.0, 10       & 1.8, 10.4, 11       & 2.5, 10.9, 12       \\ \hline
N18 2-com               & 1.0, 10.0, 10       & 1.4, 10.2, 11       & 1.8, 10.3, 12       \\ \hline
N26 3-reg                 & 1.0, 14.0, 14       & 1.8, 14.4, 15       & 2.6, 14.9, 16       \\ \hline
N26 2-com               & 1.0, 14.0, 14       & 1.6, 14.2, 15       & 2.0, 14.5, 16       \\ \hline
\end{tabular}
}
\caption{The QDCA with the METIS partitioner. For each benchmark graph type and value of $m_{cuts}$ we report the average number of cuts needed to split the circuit, the average size of the larger fragment half, and the maximum size circuit encountered.}
\label{tab:cutsAndqubits}
\end{table}

Finally, Table \ref{tab:cutsAndqubits} reports the average number of cuts and qubits as well as the maximum number of qubits needed to solve MIS on graphs containing 18 and 26 nodes. By cutting the variational ansatz into just two subcircuits, we show that the QDCA enables QPUs containing only 14 qubits to solve MIS instances on 26-node graphs, extending their reach to problem sizes up to $85\%$ larger than their own qubit count.

\subsection{Noisy simulation and evaluation on quantum hardware}\label{subsec:hardware}

\begin{figure}[t]
    \centering
    \includegraphics[width=\columnwidth]{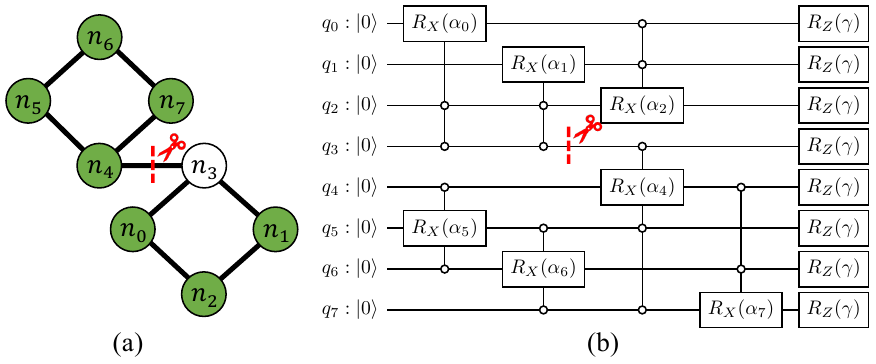}
    \caption{For our hardware experiments, \textbf{(a)} we consider an 8-node target graph partitioned into two 4-node subgraphs. Seven of the partial mixers are turned on (green nodes) while the partial mixer corresponding to $n_3$ is turned off (white node). \textbf{(b)} The corresponding QDCA ansatz which is split into 4- and 5-qubit subcircuits. The gate parameters $\{\alpha_0,\dots, \alpha_7, \gamma\}$ were obtained via classical simulation and variational optimization.}
    \label{fig:algiersansatz}
\end{figure}

\begin{figure}[t]
\centering
\includegraphics[width=\columnwidth]{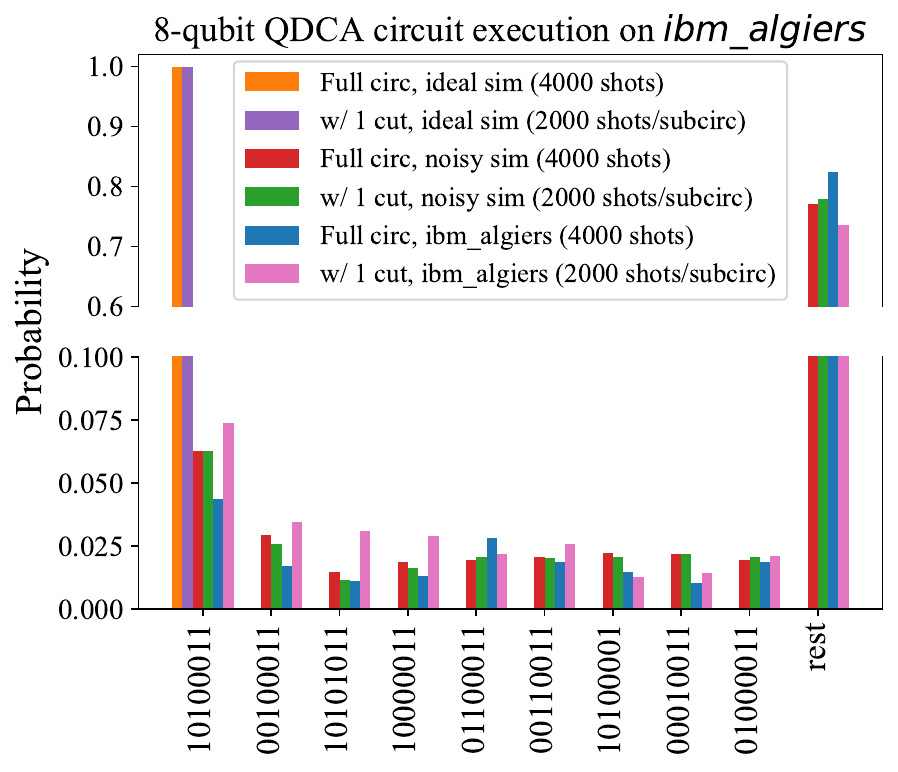}
\caption{Results of executing a 8-qubit QDCA circuit on \texttt{ibm\_algiers} with and without circuit cutting. Note the break in the Y-axis to avoid obscuring the smaller probabilities. The hardware evaluations utilizing the circuit cutting method (pink) have a higher probability of measuring the ideal \texttt{10100011} state compared to the evaluation of the full circuit (blue).}
\label{fig:algiers}
\end{figure}

We evaluated an 8-qubit QDCA ansatz corresponding to the target graph shown in Figure \ref{fig:algiersansatz} using simulations as well as the 27-qubit \texttt{ibm\_algiers} superconducting QC available via IBM Runtime \cite{davis2022runtime}.
Figure \ref{fig:algiers} shows the probability distributions that were measured after performing ideal and noiseless simulation as well as evaluation on \texttt{ibm\_algiers}. Additionally, we compare evaluations of the full, uncut ansatz against the reconstructed outputs obtained via quantum circuit cutting which splits the ansatz in Figure \ref{fig:algiersansatz}(b) with a single cut on qubit $q_3$.

To enable a fair comparison, we first optimized the ansatz via noiseless simulation --- producing an output distribution with 100\% of the weight on the \texttt{10100011} bitstring, which is in fact a solution to the MIS problem on the target graph.
Note that the bitstrings have a little-endian ordering such that they correspond to the states $\ket{q_7q_6 \dots q_0}$, and they are ordered along the X-axis in descending order; Figure \ref{fig:algiers} shows the nine highest probability bitstrings across all of the evaluation modes. The observed probabilities of the remaining bitstrings (all of which are individually smaller than the top nine shown) are summed together under ``rest''.
The optimized circuit parameters were then used for the remainder of the ideal, noisy, and hardware evaluations. 

The noisy simulations use a noise model which applies a probabilistic Pauli error channel after each gate operation. In this noise model each of the $X$, $Y$, and $Z$ error channels has a $0.5\%$ chance of occurring after every gate operation in the simulated program. The noise parameters were selected to match the performance that was seen for the hardware executions, and notably there is very little deviation between the cut and uncut output distributions in both the ideal and noisy simulations.

In the hardware evaluations, the probability of measuring the optimal MIS bitstring is substantially higher ($7.37\%$, pink bar) when utilizing quantum circuit cutting compared to the uncut circuit evaluation ($4.38\%$, blue bar).
When evaluating the QDCA on hardware, quantum circuit cutting \textit{improves} the probability of measuring the ideal outcome because it makes a tradeoff between evaluating one, large circuit and executing many, smaller circuits \cite{tang2021cutqc}.
When mapped onto \texttt{ibm\_algiers}, the full QDCA ansatz, Figure \ref{fig:algiersansatz}(b), contains 176 \texttt{CNOT}s and 344 single-qubit gates acting on eight qubits with a total circuit depth of 335.
In contrast, the two subcircuits contain (66 \texttt{CNOT}s, 115 single-qubit) and (102 \texttt{CNOT}s, 189 single-qubit) gates, act on 4 and 5 qubits, with circuit depths of 124 and 215, respectively.

\section{Related Work}\label{sec:priorwork}
The quantum circuit cutting technique has been introduced and developed by multiple prior works \cite{peng2020simulating, tang2021cutqc, perlin2021quantum}. Furthermore, prior work has employed graph partitioning algorithms as part of a divide and conquer approach to scaling the reach of the QAOA \cite{li2022large}. 
The novelty of the QDCA lies in the combination of these two approaches. 

Whereas prior work on quantum circuit cutting focused on minimizing the number of cuts needed to partition benchmark circuits such as \textit{Bernstein-Vazirani} and \textit{Supremacy} \cite{tang2021cutqc}, the QDCA applies an application-compiler co-design approach to construct the quantum circuits in a way that guarantees at most $m_{cuts}$ are needed to partition it. The QDCA incorporates the circuit cutting technique as a compilation tool within a larger application-specific framework. By targeting combinatorial optimization problems, the unwieldy exponential overheads encountered when cutting general quantum circuits can be curtailed by taking a divide and conquer approach.

The divide and conquer approach to QAOA developed in \cite{li2022large}, targeted MaxCut instances, an unconstrained combinatorial optimization problem, by partitioning the target graph into suitably small components that could be solved by a small NISQ processor or via simulation. The MaxCut solution over the original graph was then obtained by various recombination policies.
The main difference between this approach and the QDCA is the enforced independence of the subproblems. The QDCA aims to improve performance by allowing for communication between subproblems. This is enabled by quantum circuit cutting techniques in the near-term, but can be adapted to quantum teleportation protocols in the longer-term.

\section{Conclusions and Future Directions}\label{sec:conclusions}

The divide and conquer paradigm is a powerful technique for tackling difficult combinatorial optimization problems~\cite{ntaimo2005million, guo2014scaling}.
In the case of MIS --- an NP-Complete problem which can efficiently represent many other NP problems such as vertex cover and graph coloring \cite{karp1972reducibility, tarjan1977finding} --- classical divide and conquer algorithms partition the graph into multiple subgraphs, independently solve MIS on each subgraph, and then recombine the results to produce a solution over the full graph. In this classical picture, it is advantageous to minimize the number of edges that cross between the partitions because each of the subproblems are independent of each other, and no information flows between subgraphs. 

The QDCA, instead, allows information to flow between subproblems and controls the overhead incurred by that information sharing with an application-compiler co-design approach to ansatz construction \cite{tomesh2021codesign}.
Specifically, the QDCA solves MIS problems on large graphs by first partitioning the target graph into separate subgraphs, and then optimizing a variational ansatz which is defined over multiple subgraphs --- allowing quantum gate operations to be performed across subgraphs.
In the near-term, the quantum communication must be reconstructed classically --- incurring an exponential overhead in terms of classical post-processing. In the future, when large-scale quantum networks are available, the quantum communication between the subgraphs can be performed coherently with only a linear overhead in communication costs. The tunability of the QDCA helps to limit the impact of low fidelity quantum communication operations, but as these operations improve over time the QDCA can be easily adapted to take full advantage.

In our simulation results, we found that the QDCA is able to find larger independent sets than the Boppana-Halld\'orsson and classical divide and conquer algorithms. Furthermore, the QDCA enabled small QCs to target problem sizes up to $85\%$ larger than the number of qubits available on a single processor. Finally, our hardware evaluations revealed the noise \textit{mitigation} effects that circuit cutting can have in the NISQ era.

There are several open directions to consider in future work. More sophisticated partitioning algorithms can produce weighted partitions, and investigating the performance of the QDCA applied to three or more partitions is an important next step. The QDCA may also be incorporated within other hybrid algorithms like the quantum local search algorithm \cite{tomesh2022quantum} which solves MIS instances on large graphs by iteratively constructing and optimizing over smaller, local neighborhoods --- the QDCA could be employed to expand the size of these local neighborhoods.
Finally, multiple QC platforms (specifically neutral atom and trapped ion architectures) are currently developing the capability to natively execute multi-qubit gate operations \cite{graham2022multi, katz2022demonstration}. This will enable more efficient gate decompositions \cite{tomesh2022multi} of the QDCA ansatz and allow for larger-scale hardware demonstrations.
Ultimately, techniques such as the QDCA will help to harness the computational resources of distributed quantum computers, extending their reach to ever larger problem sizes.

\section*{Acknowledgements}
T.T. is supported by EPiQC, an NSF Expedition in Computing, under grant CCF-1730082. Z.S. and M.S. are supported by the National Science Foundation under Award No. 2037984, and by the the NSF Quantum Leap Challenge Institute on Hybrid Quantum Architectures and Networks (HQAN). Z.S is also supported by QNEXT. P.G. acknowledges funding by the US Department of Energy Office, Advanced Manufacturing Office (CRADA No. 2020-20099.); and by the National Science Foundation under Grant No. 2110860. This material is based upon work supported by (while was serving at) the National Science Foundation.

The submitted manuscript has been created by UChicago Argonne, LLC, Operator of Argonne National Laboratory (“Argonne”). Argonne, a U.S. Department of Energy Office of Science laboratory, is operated under Contract No. DE-AC02-06CH11357. The U.S. Government retains for itself, and others acting on its behalf, a paid-up nonexclusive, irrevocable worldwide license in said article to reproduce, prepare derivative works, distribute copies to the public, and perform publicly and display publicly, by or on behalf of the Government. The Department of Energy will provide public access to these results of federally sponsored research in accordance with the DOE Public Access Plan. \url{http://energy.gov/downloads/doe-public- access-plan}.


\bibliographystyle{IEEEtran}
\bibliography{main}

\end{document}